\documentclass[12pt]{article}
\usepackage{graphicx}
\usepackage{epstopdf}
\usepackage{amsthm}
\usepackage{amsmath}
\usepackage{amssymb}
\usepackage{amstext}
\usepackage{amsfonts}
\usepackage{enumerate}
\usepackage[authoryear,nonamebreak,bibstyle]{natbib}
\usepackage{footnote}
\makesavenoteenv{tabular}
\makesavenoteenv{table}
\usepackage[onehalfspacing]{setspace}
\usepackage[shortlabels]{enumitem}
\usepackage{multirow}
\usepackage[title]{appendix}
\usepackage{xcolor}
\usepackage{cancel}
\usepackage[normalem]{ulem}
\usepackage{verbatim}

\begin{document}
\baselineskip=.22in\parindent=30pt

\newtheorem{tm}{Theorem}
\newtheorem{dfn}{Definition}
\newtheorem{lma}{Lemma}
\newtheorem{assu}{Assumption}
\newtheorem{prop}{Proposition}
\newtheorem{cro}{Corollary}
\newtheorem*{theorem*}{Theorem}
\newtheorem{example}{Example}
\newtheorem{observation}{Observation}
\newtheorem{conjecture}{Conjecture}
\newcommand{\exm}{\begin{example}}
\newcommand{\exmm}{\end{example}}
\newcommand{\obs}{\begin{observation}}
\newcommand{\obss}{\end{observation}}
\newcommand{\con}{\begin{conjecture}}
\newcommand{\conn}{\end{conjecture}}
\newcommand{\cor}{\begin{cro}}
\newcommand{\corr}{\end{cro}}
\newtheorem{exa}{Example}
\newcommand{\ex}{\begin{exa}}
\newcommand{\exx}{\end{exa}}
\newtheorem{remak}{Remark}
\newcommand{\rmk}{\begin{remak}}
\newcommand{\rmkk}{\end{remak}}
\newtheorem{nremak}{\it Remark}
\newcommand{\nrmk}{\begin{nremak}}
\newcommand{\nrmkk}{\end{nremak}}
\newcommand{\thm}{\begin{tm}}
\newcommand{\nt}{\noindent}
\newcommand{\thmm}{\end{tm}}
\newcommand{\lm}{\begin{lma}}
\newcommand{\lmm}{\end{lma}}
\newcommand{\ass}{\begin{assu}}
\newcommand{\asss}{\end{assu}}
\newcommand{\df}{\begin{dfn}  }
\newcommand{\dff}{\end{dfn}}
\newcommand{\prp}{\begin{prop}}
\newcommand{\prpp}{\end{prop}}
\newcommand{\bqu}{\sloppy \small \begin{quote}}
\newcommand{\equ}{\end{quote} \sloppy \large}
\newcommand\cites[1]{\citeauthor{#1}'s\ (\citeyear{#1})}

\newcommand{\eq}{\begin{equation}}
\newcommand{\eqq}{\end{equation}}
\newtheorem{claim}{\it Claim}
\newcommand{\cl}{\begin{claim}}
\newcommand{\cll}{\end{claim}}
\newcommand{\bit}{\begin{itemize}}
\newcommand{\eit}{\end{itemize}}
\newcommand{\ben}{\begin{enumerate}}
\newcommand{\een}{\end{enumerate}}
\newcommand{\bcen}{\begin{center}}
\newcommand{\ecen}{\end{center}}
\newcommand{\fn}{\footnote}
\newcommand{\ds}{\begin{description}}
\newcommand{\dss}{\end{description}}
\newcommand{\prf}{\begin{proof}}
\newcommand{\prff}{\end{proof}}
\newcommand{\cs}{\begin{cases}}
\newcommand{\css}{\end{cases}}
\newcommand{\ml}{\item}
\newcommand{\lb}{\label}
\newcommand{\ra}{\rightarrow}
\newcommand{\tra}{\twoheadrightarrow}
\newcommand*{\supp}{\operatornamewithlimits{sup}\limits}
\newcommand*{\inff}{\operatornamewithlimits{inf}\limits}
\newcommand{\nf}{\normalfont}
\renewcommand{\Re}{\mathbb{R}}
\newcommand*{\mmax}{\operatornamewithlimits{max}\limits}
\newcommand*{\mmin}{\operatornamewithlimits{min}\limits}
\newcommand*{\argmax}{\operatornamewithlimits{arg max}\limits}
\newcommand*{\argmin}{\operatornamewithlimits{arg min}\limits}
\newcommand{\uhr}{\!\! \upharpoonright  \!\! }

\newcommand{\CR}{\mathcal R}
\newcommand{\CC}{\mathcal C}
\newcommand{\CT}{\mathcal T}
\newcommand{\CS}{\mathcal S}
\newcommand{\CM}{\mathcal M}
\newcommand{\CL}{\mathcal L}
\newcommand{\CP}{\mathcal P}

\newtheorem{innercustomthm}{Theorem}
\newenvironment{customthm}[1]
  {\renewcommand\theinnercustomthm{#1}\innercustomthm}
  {\endinnercustomthm}
\newtheorem{einnercustomthm}{Extended Theorem}
\newenvironment{ecustomthm}[1]
  {\renewcommand\theeinnercustomthm{#1}\einnercustomthm}
  {\endeinnercustomthm}
  
  \newtheorem{innercustomcor}{Corollary}
\newenvironment{customcor}[1]
  {\renewcommand\theinnercustomcor{#1}\innercustomcor}
  {\endinnercustomcor}
\newtheorem{einnercustomcor}{Extended Theorem}
\newenvironment{ecustomcor}[1]
  {\renewcommand\theeinnercustomcor{#1}\einnercustomcor}
  {\endeinnercustomcor}
    \newtheorem{innercustomlm}{Lemma}
\newenvironment{customlm}[1]
  {\renewcommand\theinnercustomlm{#1}\innercustomlm}
  {\endinnercustomlm}

\newcommand{\red}{\textcolor{red}}
\newcommand{\blue}{\textcolor{blue}}
\newcommand{\purple}{\textcolor{purple}}
\newcommand{\mred}[1]{\color{red}{#1}\color{black}}
\newcommand{\mblue}[1]{\color{blue}{#1}\color{black}}
\newcommand{\mpurple}[1]{\color{purple}{#1}\color{black}}

\makeatletter
\newcommand{\customlabel}[2]{%
\protected@write \@auxout {}{\string \newlabel {#1}{{#2}{}}}}
\makeatother


\def\qed{\hfill\vrule height4pt width4pt
depth0pt}
\def\reff #1\par{\noindent\hangindent =\parindent
\hangafter =1 #1\par}
\def\title #1{\begin{center}
{\Large {\bf #1}}
\end{center}}
\def\author #1{\begin{center} {\large #1}
\end{center}}
\def\date #1{\centerline {\large #1}}
\def\place #1{\begin{center}{\large #1}
\end{center}}

\def\date #1{\centerline {\large #1}}
\def\place #1{\begin{center}{\large #1}\end{center}}
\def\intr #1{\stackrel {\circ}{#1}}
\def\R{{\rm I\kern-1.7pt R}}
 \def\N{{\rm I}\hskip-.13em{\rm N}}
 \newcommand{\cprod}{\Pi_{i=1}^\ell}
\let\Large=\large
\let\large=\normalsize


\begin{titlepage}
\def\thefootnote{\fnsymbol{footnote}}
\vspace*{0.5in}

\title{On an Extension of a Theorem of Eilenberg \\ \vspace{0.2em} 
and a Characterization of Topological Connectedness\fn{The two theorems reported here were announced without proof  at the {\it 2019 NSF/NBER/CEME Conference on Mathematical Economics:  In Honor of Robert M.  Anderson} in Berkeley, October 25, 2019; and more comprehensively,  at a departmental seminar at the {\it Australian National University} on December 17, 2019.  The authors  acknowledge with gratitude extended correspondence  and conversation with Professors Max Amarante, Yorgos Gerasimou, Alfio Giarlotta, Farhad Husseinov, Michael Mandler, Rich McLean and Debraj Ray. Needless to say, all errors of reading and interpretation are solely the authors'. We use the {\it certified random order} in order to list the authors; see Ray-Robson (2018, American Economic Review).    \vspace{0.05em}  }}

\vskip 0.5em

\author{M. Ali Khan\fn{Department of Economics, Johns Hopkins University, Baltimore, MD 21218. {\bf Email:} akhan@jhu.edu.   
 }  \textcircled{r} 
 Metin Uyan{\i}k\footnote{School of Economics, The University of Queensland, Brisbane, QLD 4072. {\bf Email:}  m.uyanik@uq.edu.au ORCID: 0000-0003-0224-7851
}}


\vskip 1.00em
\date{December 24, 2019}

\vskip 1.75em

\vskip 1.00em

\baselineskip=.18in

\noindent{\bf Abstract:} 
On taking a non-trivial and semi-transitive bi-relation constituted by two    ({\it hard} and~{\it soft}) binary relations,   we report a  (i) $p$-continuity  assumption that guarantees the completeness and transitivity of its {\it soft}  part, and a  (ii) characterization of a connected topological space in terms of its attendant  properties on the space. Our work generalizes  antecedent results in applied mathematics, all following  \cite{ei41}, and now framed in the context of a parametrized-topological space. This re-framing is   directly inspired by the continuity assumption in \cite{wo43} and the mixture-space structure proposed in \cite{hm53}, and the  unifying synthesis of these  pioneering but neglected papers  that it affords  may have independent interest. 
 
 \vskip 0.25in


%
%
%


\medskip


%
%

\end{titlepage}

\setcounter{footnote}{0}


\setlength{\abovedisplayskip}{1pt}
\setlength{\belowdisplayskip}{1pt}


\newif\ifall
\alltrue 

%
%
%
%
%
%
%



\section{Introduction} 


The question considered in this paper concerns a  binary relation  $R$ on a set $X$ conceived as a subset of $X\times X,$  with its 
{\it transpose}, and {\it upper} and  {\it lower  section} at $x\in X$ respectively defined as 
$ R^{-1} = \{(x,y)~|~(y,x)\in R\},\;  R(x) = \{y~|~ (x,y)\in R\},$ and $R^{-1}(x) = \{y~|~ (y,x)\in R\}$.  With $\Delta=\{(x,x)|x\in X\}, \;R^c$  the complement of $R,$  its  {\it symmetric}  and  {\it asymmetric} parts  respectively denoted as  $I = R\cap R^{-1}, \mbox{ and } P = R\backslash R^{-1}, \; I \cap P = emptyset,$  and $R=I\cup P,$  we let  the {\it composition} $R\circ R'$ be given by  $(y,x)\in R\circ R'$ if  $R^{-1}(x)\cap R'(y)\neq\emptyset,$ for any two relations $R, R'$ on a  $X$, and  call a relation       $R$ on $X$   
{\it non-trivial} if  $P\neq\emptyset$, {\it semi-transitive} if  $I\circ P \subseteq P$ and $P\circ I \subseteq P$,   
{\it transitive} if $R\circ R\subseteq R$,   
({\it negatively transitive} if $R^c$ is transitive),  
 and  {\it complete} if  $R\cup R^{-1} = X\times X$.  
 We can now ask  for a sufficient condition in any register that ensures that a non-trivial, semi-transitive relation with a transitive symmetric part is complete  and transitive. Thus, rather than denoting $R$  by $\preccurlyeq$, as is standard especially in the social sciences,\fn{This notation is perhaps original to \cite[p. 11]{ar51} who observed that he was representing  \lq\lq preference [relations] by a notation not customarily employed in economics, though familiar in mathematics and particularly in symbolic logic." } this is to ask for any register that ensures,  in  the vernacular of 
set-theory,  
\vspace{-12pt}

$$
(P \!\neq \!\emptyset) \land (\!\!\!~(I\circ P \!\subseteq \! P) \land (P\circ I \!\subseteq \! P) \! \!\!~) \land  (I\circ I \!\subseteq \!  I)\! \Longrightarrow\! (R\cup R^{-1} \! = \! X \! \times \! X) \land (R\circ R \!\subseteq \! R).
$$ 
\vspace{-12pt}

 In his remarkable paper,  \cite{ei41} provided a partial answer this question by invoking  a topological register and assuming  a continuous binary (preference) relation on a connected (choice) set. His answer inaugurated the study of  a partially ordered topological space, and his work   received important substantive extension at the hands of \cite{de54, de64},  Ward (\citeyear{wa54b} and  \citeyear{wa54}), \cite{so65, so67}, \cite{mc66}  and \cite{sc71}.  In the twin-registers of economic  and   decision theory, the extension involved a move to a setting of total pre-orders, which is to say from a setting of singleton indifference sets to a more general situation where  the preference map of a consumer is  delineated by indifference surfaces.

  Recent work in both mathematical economics and mathematical psychology  has rediscovered these original papers, and under the label of  the  Eilenberg-Sonnenschein (ES) research program, has been particularly stimulated by what it sees as the derivation of  behavioral consequences of  merely  technical topological assumptions; see,   for example,   \cite{ku19a},  \cite{gw19}  and their references.  Eilenberg, and following him Ward,  phrased their  results only in the language of topological structures, but  \cite{hm53},  with \cite{wo43} as their important precursor, focused on the  functional representation of the relation, and  
 shifted  all topological assumptions on the choice  set to that on the unit interval. This is  to say, they  focused their attention  on the mixing operation, rather than on the objects of choice itself.  \cite{wo43} used a similar scalar-continuity property in  his  work, but with an additional  monotonicity assumption; also see \cite{fi82}, \cite{wa89}, \cite{bm95}, \cite{hp01} and \cite*{cim12}, \citet*{gku19} on numerical representation of preferences. In addition to this parametrized topological setting, we also mention for the record, a   third setting, that involves no topology at all, but constrains itself to a   purely algebraic structure. This goes back at least to seminal paper of 
 Holder (1901); also see \cite{fi72} and \cite{lu00}.

In this paper we work with a richer order-theoretic structure defined by two binary relations instead of one, a structure that is only now being appreciated and given prominence;  see \cite{gg13}, \cite{gi14}, \cite{gw19} and \cite{uk19a} and their references; also see \citet{ch71} and \cite*{mmr06}  for an implicitly assumed bi-preference structure.    In this richer structure, we introduce a parametric continuity concept for (bi-)relations that does not assume any structure on the choice set itself,  and is weaker than the usual  continuity properties that require it  to have one of the topological or algebraic structures. We report two theorems, one of which provides a new characterization of topological connectedness, and show by examples that our results are non-vacuous.    Even in the special case of a single binary relation, our first result provide a synthetic treatment of  the antecedent ES literature  by generalizing and unifying it, and in particular, our analytical treatment  provides an alternative proof of the theorems of \cite{ei41}, \cite{so65, so67}  and \cite{ra63}.  
 
%


\section{Notational and Conceptual Preliminaries } \lb{sec: mp}

We elaborate the terminology presented in Section 1 to the setting of a bi-relation. 


%
\vspace{-1pt}

\df
A bi-relation on a set $X$ is a pair $(R_H, R_S)$ of relations on $X$ such that $R_H\subseteq R_S$ and $P_S\subseteq P_H$. 
A bi-relation $(R_H, R_S)$ is {\it non-trivial} if $P_S\neq\emptyset$, and   {\it semi-transitive} if 
\ben[{\nf (i)}, topsep=-1pt]
\setlength{\itemsep}{-1pt} 
\ml $R_H$ is semi-transitive, 
\ml $I_S\circ R_H\subseteq R_S$ and $R_H\circ I_S\subseteq R_S$,
\ml $P_H\cap \left( P_S\circ I_S\right)\subseteq P_S$  and $P_H\cap \left(I_S \circ P_S \right)\subseteq P_S$.
\een
\lb{dfn: bp} 
\dff 

\nt For any bi-relation $(R_H, R_S)$, we informally refer to $R_H$ as its {\it hard} part,  and $R_S$ as its {\it soft} part. We now present the basic continuity assumption that motivates this paper and is the subject of its investigation.  

\df
A bi-relation $(R_H, R_S)$ on a set $X$ is  $p$-continuous (parametrically continuous)  if for all $x,y\in X$, there exist a topological space $\Lambda_{xy}$ and a function $f_{xy}:\Lambda_{xy}\ra X$  with $x,y\in f_{xy}(\Lambda_{xy})$ such that for all $z\in X$, 
\ben[{\nf (i)}, topsep=-1pt]
\setlength{\itemsep}{-1pt} 
\ml  $f_{xy}^{-1}(R_H(z))$ and $f_{xy}^{-1}(R_H^{-1}(z))$ are closed,  and
\ml $f_{xy}^{-1}(P_S(z))$ and $f_{xy}^{-1}(P_S^{-1}(z))$ are open. 
\een
\lb{df: bicontinuous}
\dff

We call a $p$-continuous bi-relation {\it connected} if each space $\Lambda_{xy}$ in Definition \ref{df: bicontinuous} is connected. The concept is admittedly abstract, but one can get a feel for it by dissociating the two aspects of \cite{hm53} that it generalizes: first, keeping the unit interval, and focusing on the fact that rather than a linear function,  any function is being considered; and second, replacing the unit interval by any topological space, both  operations localized by the dependence on the pair $(\Lambda_{xy}, f_{xy})$  on the two chosen points, $x$ and $y$.  

We now recall for the reader the standard 

\df
A topological space $X$ is {\it connected} if it is not the union of two non-empty, disjoint open sets.   
A subset of $X$ is {\it connected} if it is connected as a subspace.  
\lb{df: connected}
\dff
\vspace{-3pt}

\nt A relation $R$ on a set $X$ is {\it $p$-continuous} if the bi-relation $(R,R)$ is $p$-continuous, and it is  {\it connected}  if the bi-relation $(R,R)$ is connected. The reader may wish to contrast this definition to \citet[p. 27]{fi72}. 
Moreover, 
\vspace{-3pt}

\df
A relation on a topological space is {\it continuous} if it has closed sections and its asymmetric part has open sections. 
\lb{dfn: seccont}
\dff

\vspace{-12pt}


\section{The Results and their Proofs} 

We can now present our first result, 

\vspace{-3pt}

\thm
The soft part of every non-trivial, semi-transitive and connected  bi-relation  is complete and transitive. 
\lb{thm: bict}
\thmm

\vspace{-3pt}

Next, we  present a special case of Theorem \ref{thm: bict} for a binary relation.  Note that a relation $R$ on a set $X$ is non-trivial, semi-transitive, connected and has a transitive symmetric part if and only if the bi-relation $(R,R)$ on $X$ is non-trivial, semi-transitivity and connected.   
%
On setting $R_H=R_S$ in a bi-relation $(R_H, R_S),$ we obtain  the following as a corollary of Theorem \ref{thm: bict}.  

\cor
Every non-trivial, semi-transitive and connected binary relation whose symmetric part is transitive, is complete and transitive. 
\lb{thm: icont}
\corr

In the light of the above discussion, we can now present 
 
  \begin{customthm}{(Eilenberg 1941)}[{}]
Every anti-symmetric, complete and continuous relation on a connected set is transitive.
\lb{thm: eil}
\end{customthm}

\nt Theorem \ref{thm: bict} provides  a generalization and alternative proof not only of  \citet[2.1]{ei41}, but also of the  extensions pursued in  \cite{ra63} and \cite{so65}.   Without any attempt to downplay the importance of the exercise,  we invite the reader to show    that the principal results in the literature follow as a consequence of Theorem \ref{thm: bict} and  Corollary \ref{thm: icont} above, and Lemma \ref{thm: sen}  and Propositions \ref{thm: seccont}--\ref{thm: sccont} below:  \citet[Lemma]{ra63}, \citet[Theorem 3]{so65}, \citet[Theorem]{sc71}, \citet[Theorem 1]{du11}, \citet[Theorem 1]{ks15}, \citet[Theorem 1]{mm18},  \citet[Proposition 1]{ku19a};    \citet[Theorem 5.2]{gw19} and finally \citet[Theorems 1 and 2]{uk19b}. In our judgment, this exercise testifies to the importance of the results presented here. (Note, however,  that the authors of the above papers present their results in an extended form that also involve consideration not germane to those pursued here;  our generalization concerns only the relevant part of their results.)   In the sequel, we also illustrate the novelty of our proof-techniques by contrasting them with earlier proofs. 

 We now turn to the proofs, and  begin by recalling  the following.

\vspace{-5pt}

\lm
For any binary relation  $R$ on a set $X$ the following is true: 
\ben[{\nf (a)}, topsep=-1pt]
\setlength{\itemsep}{-1pt} 
\ml if $R$ is complete and semi-transitive,  then $I$ is transitive;
\ml if $P$ is negatively transitive, then it is transitive and $R$ is semi-transitive; 
\ml $R$ is transitive if and only if it is  semi-transitive and $P, I$ are transitive.
\een
\lb{thm: sen}
\lmm

\vspace{-3pt}

\nt For a proof, see  \citet[Theorem I]{se69} and \citet[Proposition 2]{ku19a} on Sen's  deconstruction of the transitivity postulate.  

We shall also need the following two claims concerning  a non-trivial, semi-transitive and connected bi-relation  $(R_H, R_S)$ on an arbitrary  set $X$.

\vspace{-5pt}

\cl
If $(y,x)\in P_S$, then $P_S(y)\cup P_S^{-1}(x)=X$. 
\lb{thm: bint}
\cll
\vspace{-12pt}

\cl
$R_S\cup R_S^{-1}=X\times X$. 
\lb{thm: bicomplete}
\cll
\vspace{-5pt}

The proof of Theorem \ref{thm: bict} follows as a direct consequence of these preliminary results 

\prf[Proof of Theorem \ref{thm: bict}]   Claim 2 already establishes the completeness of the soft part of the bi-relation. Now negative transitivity of $P_S$ is equivalent to Claim \ref{thm: bint}. Then  (b) of Lemma \ref{thm: sen} implies  $R_S$ is semi-transitive and $P_S$ is transitive.  It follows from Claim \ref{thm: bicomplete}, semi-transitivity of $R_S$  and  (a) of Lemma \ref{thm: sen} that $I_S$ is transitive. Then (c) of Lemma \ref{thm: sen} implies $R_S$ is transitive.    \prff

All that remains now are the proofs of the two claims. 

\prf[Proof of Claim \ref{thm: bint}]
Pick $(y,x)\in P_S$.  Then $(y,x)\in P_H$. It follows from semi-transitivity of the bi-relation that
\eq
R_H(y)\cup R_H^{-1}(x)=P_S(y)\cup P_S^{-1}(x).
\lb{eqn: cup}
\eqq
One direction of the inclusion relationship immediately follows from the definition of bi-relation. In order to prove the other direction pick  $z\in R_H^{-1}(x)$.  Then either $z\in I_H(x)$ or $z\in P_H^{-1}(x)$. Assume $z\in I_H(x)$.  It follows from the definition of bi-relation that $z\in I_S(x)$. 
Moreover, $x\in P_H(y)$, $z\in I_H(x)$  and semi-transitivity of $R_H$ imply $(y,z)\in P_H$.   
Note that $z\in I_S (x)$ and $x\in P_S(y)$ imply $(y,z)\in I_S\circ P_S$.  Then semi-transitivity of the bi-relation implies  $z\in P_S(y)$.  
Now assume $z\in P_H^{-1}(x)$. Then the definition of bi-relation implies that either $x\in P_S(z)$ or $x\in I_S(z)$. If $x\in P_S(z)$, then there is nothing to prove.   
Now assume $x\in I_S(z)$.  Then semi-transitivity of the bi-relation, $z \in I_S(x)$ and $x\in P_H(y)$ imply that $z\in R_S(y)$. Hence either $z \in P_S(y)$ or $z\in I_S(y)$. If $z\in I_S(y)$, then semi-transitivity of the bi-relation, $x\in P_S(y), y\in I_S(z)$ and $x\in P_H(z)$ imply that $x\in P_S(z)$. This contradicts $x\in I_S(z)$. Hence, $z \in P_S(y)$ must hold.  
 The proof for  $z\in R_H(y)$ is analogous.

We next prove that $P_S(y)\cup P_S^{-1}(x)=X$. To this end pick $z\in X$. It follows from the connectedness of the bi-relation that there exist a connected topological space $\Lambda_{yz}$ and a function $f_{yz}:\Lambda_{yz} \ra X$ satisfying the conditions in Definition \ref{df: bicontinuous}. It follows from Equation \ref{eqn: cup} above that 
\eq
f_{yz}^{-1}(R_H(y))\cup f_{yz}^{-1}(R_H^{-1}(x))
=f_{yz}^{-1}P_S(y)\cup f_{yz}^{-1}(P_S^{-1}(x)).
\lb{eqn: preimagecup}
\eqq
It follows from $x\in P_S(y)$ and $y=f_{yz}(\lambda)$ for some $\lambda\in \Lambda_{yz}$ that $\lambda\in f_{yz}^{-1}(P_S^{-1}(x))$. Hence the set in Equation \ref{eqn: preimagecup} is non-empty. It follows from $p$-continuity that it is both closed and open in $\Lambda_{yz}$. Therefore, as a nonempty, closed and open subset of a connected set $\Lambda_{yz}$,  the set $f_{yz}^{-1}(P_S(y))\cup f_{yz}^{-1}(P_S^{-1}(x))$ is equal to $\Lambda_{yz}$.  Then it follows from $z=f_{yz}(\delta)$ for some $\delta\in \Lambda_{yz}$ that   $\delta\in f_{yz}^{-1}(P_S(y))\cup f_{yz}^{-1}(P_S^{-1}(x))$, hence $z\in P_S(y)$ or $z\in P_S^{-1}(x)$.  
\prff

\vspace{-8pt}

\prf[Proof of Claim \ref{thm: bicomplete}]
Assume there exists $u,v\in X$ such that $(u, v)\notin R_S\cup R_S^{-1}$. Note that non-triviality of $R_S$ implies that $(\bar y, \bar x)\in P_S$ for some $\bar x,\bar y\in X$.  Since $P_S$ is negatively transitive, then $u\in P_S^{-1}(\bar x)$ or $u\in P_S(\bar y)$. Assume $u\in P_S^{-1}(\bar x)$. 	Then negative transitivity of $P_S$ implies that $v\in P_S^{-1}(\bar x)$ or $v\in P_S(u)$. Since $u,v\notin R_S\cup R_S^{-1}$, therefore  $v\in P_S^{-1}(\bar x)$. Hence $\bar x\in P_S(u)\cap P_S(v)$. The semi-transitivity of the bi-relation implies that  
\eq
I_H(u)\cap R_H(v)= \emptyset=R_H(u)\cap I_H(v).
\lb{eqn: icap}
\eqq
In order to prove the first equality, assume there exists $z\in I_H(u)\cap R_H(v)$.    If $z\in P_H(v)$, then semi-transitivity of $R_H$ implies that $u\in P_H(v)$, hence $u\in R_S(v)$. This yields a contradiction. Then assume $z\in I_H(v)$. The definition of bi-relation implies that  $z\in I_S(v)$. 
Then it follows from semi-transitivity of bi-relation that $(u,v)\in R_S$. This yields a  contradiction. The proof of the second equality is analogous. 

Therefore, Equation \ref{eqn: icap} and the definition of the bi-relation imply that  
\eq
R_H(u)\cap R_H(v)=P_S(u)\cap P_S(v).
\lb{eqn: cap}
\eqq
It follows from the connectedness of the bi-relation  that there exist a connected topological space $\Lambda_{\bar x u}$ and a function $f_{\bar x u}:\Lambda_{\bar x u}\ra X$ satisfying the conditions in Definition \ref{df: bicontinuous}. It follows from Equation \ref{eqn: cap} above that  
\eq
f_{\bar x u}^{-1}(R_H(u))\cap f_{\bar x u}^{-1}(R_H(v))
=f_{\bar x u}^{-1}(P_S(u))\cap f_{\bar x u}^{-1}(P_S(v)).
\lb{eqn: preimagecap}
\eqq
It follows from $\bar x\in P_S(u)\cap P_S(v)$ and $\bar x=f_{\bar x u}(\lambda)$ for some $\lambda\in \Lambda_{\bar x u}$ that $\lambda\in f_{\bar x u}^{-1}(P_S(u))\cap f_{\bar x u}^{-1}(P_S(v))$. Analogously it follows from $u\notin P_S(u)$ and $u=f_{\bar x u}(\delta)$ for some $\delta\in \Lambda_{\bar x u}$ that $\delta\notin f_{\bar x u}^{-1}(P_S(u))\cap f_{\bar x u}^{-1}(P_S(v))$. Hence the set if Equation \ref{eqn: preimagecup} is non-empty and proper subset of $\Lambda_{yz}$. It follows from $p$-continuity that it is both closed and open in $\Lambda_{yz}$. This contradicts with the connectedness of $\Lambda_{\bar x u}$. 

The proof is analogous for $u\in P_S(\bar y)$.   Therefore $R_S$ is complete.\prff




We now turn to the second main result of this paper. Note that the essential point of   Theorem \ref{thm: bict} is that its conclusion of completeness and transitivity of the soft relation  do not call  for {\it any} mathematical structure on the (choice) set on which the (preference)  bi-relation is defined: a local topological structure on the local  parameter space suffices enough to obtain.  
We now work towards a converse question first posed in the context of a single binary relation in  \cite{ku19a}. Our second theorem shows that topological connectedness is both necessary and sufficient for completeness and transitivity of a bi-relation.  
 Towards this end, for any topology $\tau$ on $X$, let $\mathcal R(X, \tau),$ denote the set of all non-trivial and semi-transitive bi-relations $(R_H,  R_S)$ on $X$ such that $\tau(R_H, R_S)\subseteq \tau$,  where $\tau(R_H, R_S)$  denotes the coarsest topology on $X$ containing all sections of $P_S$ and the complements of the sections of $R_H$.  We can then present

\thm
For any set $X$ and topology $\tau$ on it the following is equivalent:   
\ben[{\nf (a)}, topsep=-1pt]
\setlength{\itemsep}{-1pt} 
\ml the space $(X,\tau)$ is connected;
\ml  if $(R, R)\in \mathcal R(X, \tau)$, then $R$ is complete and transitive;  
\ml  if $(R_H, R_S)\in \mathcal R(X, \tau)$, then $R_S$ is complete and transitive. 
\een
\lb{thm: topology}
\thmm

\vspace{-5pt}  

\prf  The implication (a)~$\Rightarrow$~(c) follows from Theorem \ref{thm: bict} above and Proposition \ref{thm: seccont} below, (c)~$\Rightarrow$~(b) from setting $R_H=R_S$ and  (b)~$\Rightarrow$~(a) from  \citet[Theorem 2]{ku19a}.~\prff

  We note that in a parallel but different path, \citet{mc66}  provides a characterization of compactness and Hausdorff separation axiom by using the preferences defined on a choice set.


We have already observed above that  the richer bi-preference structure that we work with enables  different techniques of proof.  We elaborate this observation here, beginning  with  {\it Eilenberg's method}. As a preliminary  remark, note that  although Eilenberg assumes the relation to be complete, what follows shows that his method-of-proof extends to a situation where this is not so.  Given his  assumption on  $R$ being an anti-symmetric and continuous relation on a connected set $X, \; R\cap R^{-1}\subseteq \Delta$.  The method requires one to pick $x,y,z\in X$ such that $(z,y),(y,x)\in P$.  The connectedness of $X$ and the continuity of $R$ imply that $R(y)$ is connected and contains $y$.  Note first  that if $y\notin R(y)$, then $x\in R(y), z\notin R(y)$ and continuity of $R$ imply that $R(y)$ is a closed, open, non-empty and a proper subset of $X$, which contradicts the connectedness of the space.  Now assume $R(y)$ is disconnected. Then there exists non-empty, disjoint and closed subsets $A, B$ of the subspace $R(y)$ such that $A\cup B=R(y)$. Let $y\in B$. Note that $X=A\cup (B\cup R^{-1}(y)\cup (R(y)\cup R^{-1}(y))^c\cup \{y\})=A\cup (B\cup R^{-1}(y)\cup (P(y)\cup P^{-1}(y))^c)$, 	where $A$ and the set in paranthesis are non-empty, disjoint and closed. This contradicts the connectedness of $X$. Finally,   note that $R(y)=P(z)\cup P^{-1}(z)\cup (R\cup R^{-1})^c(z),$ and recall that $x,y\in R(y)$ and $y\in P(z)$. If $x\notin P(z)$, then this furnishes us a contradiction to the connectedness of $R(y)$. Therefore, $(z,x)\in P$.  Since $R$ is anti-symmetric, it is transitive.    The proof is complete.  

Now consider the special case of the method we pursue in Theorem \ref{thm: bict} which begins with   an anti-symmetric and continuous relation  $R$ on a connected set $X$, and picks  $x,y\in X$ such that $(y,x)\in P$.  Then the anti-symmetry of $R$ implies that $R(y)\cup R^{-1}(x)= P(y)\cup P^{-1}(x)$. Therefore it follows from the continuity of $R$ and connectedness of $X$ that $P(y)\cup P^{-1}(x)=X$. Hence $P$ is negatively transitive, which implies $P$ is transitive. In order to see this,   
pick $(y,z), (z,x)\in P$. Negative transitivity of $P$ implies that $(y,x)\in P$ or $(z,y)\in P$. Therefore $(y,x)\in P$, hence $P$ is transitive. Since $R$ is anti-symmetric, it is transitive. Therefore the proof is complete.

Finally, we leave it to the reader to check that our method of proof has some similarities with the argument pursued by \cite{sc71},  but there are also   some subtle differences. In  any case, is not the intricacy of the proofs but the surprise of the conclusion that is the point of all this work.       

We conclude  this section by listing possible future directions of research. Note that $p$-continuity does not require $f_{xy}$ to be mixture-linear, and more importantly, for all $x,y$ the function is allowed to be chosen differently. This has relevance to  
\citet*{daz81} which studies different forms of convexity assumptions on functions, and merits investigation. It will also  be interesting to extend the results in this paper to $k$-connected spaces and component-wise non-trivial relations on an arbitrary topological spaces; see for example \cite{ku19a}. Furthermore,  given our  tilt to the subject, generalizing  \cites{ lo67} set-theoretic version of Eilenberg and Sonnenschein's result is of interest. Finally, it will be interesting to explore how our results generalize to  $n$-ary relations, given their importance in the literature of analytical philosophy; see \cite{an93,te96,te15} and their references.  

\vspace{-8pt}

%
%
%
%

\section{Some Additional Considerations }

In this section, we turn to   the usual scalar and section continuity properties
of a relation, and show that they are stronger than  $p$-continuity, once relevant mathematical structures on the choice set are in place;  see \cite{uk19a} for a detailed discussion of the relationship between different continuity assumptions on preferences, and  \cite{cm16} for those on functions. 
\vspace{-5pt}

\prp
Every continuous relation on a topological space is $p$-continuous. 
\lb{thm: seccont}
\prpp
%

\prf  Let $R$ be a continuous relation on a topological space $X$. 
 For all $x,y\in X$, setting $\Lambda_{xy}=X$ and $f_{xy}(a)=a$ for all $a\in \Lambda_{xy}$ finishes the proof. Finally, note that if $X$ is connected, then the bi-relation is  connected.  \prff

For scalar continuity concepts, we say that  a set $\CS$ is a {\it mixture set} if for any $x,y\in \CS$ and for any $\mu\in [0,1]$ we can associate another element,  $x \mu y,$ that is  in $\CS,$ and where for  all $\lambda,\mu\in [0,1]$,$$   \mbox{(S1)} \; x1y = x,   \mbox{(S1)}  x\mu y =  y(1-\mu)x,  \mbox{and (S3)} (x \mu y)\lambda  y =  x (\lambda\mu) y.$$

 \vspace{-1pt}  
 
\df
We refer a relation  $R$ on a  mixture set $\CS$  to be  
\ben[{\nf (i)}, topsep=0pt]
\setlength{\itemsep}{0pt} 
\ml mixture-continuous if for all $x,y,z\in \CS,$ the sets $\{\lambda\in [0,1]~|~x\lambda y \in R(z)\}$ and $\{\lambda \in [0,1] ~|~x\lambda y\in R^{-1}(z)\}$ are closed, 

\ml Archimedean if for all $x,y,z\in\CS$ with $(y,x)\in P$, there exists $\lambda, \delta \in (0,1)$  such that $x\lambda z \in P(y)$ and   $y\delta z\in P^{-1}(x)$.
\een
Moreover, $R$ is scalarly continuous if it is mixture-continuous and Archimedean. 

\lb{df: sccont}
\dff 
\vspace{-8pt}  

\prp
Every scalarly continuous relation on a mixture set is connected, hence $p$-continuous. 
\lb{thm: sccont}
\prpp

\vspace{-12pt}

\prf   Let $R$ be a scalarly continuous relation on a mixture set $X$  and $\Lambda_{xy}=[0,1]$ for all $x,y\in X$. Then, for all  $x,y\in X$, $f_{xy}(\lambda)=x\lambda y$ for all $\lambda\in \Lambda_{xy}$.   It is easy to observe that the closedness of the sets $f_{xy}^{-1}(R(z))$ and $f_{xy}^{-1}(R^{-1}(z))$ is equivalent to mixture-continuity property. Under the mixture continuity assumption, it follows from \citet*[Proposition 1]{gku19} that Archimedean property is equivalent to the property that $f_{xy}^{-1}(P(z))$ and $f_{xy}^{-1}(P^{-1}(z))$ are open. 
\prff  


The following two definitions extends the usual continuity assumptions on uni-relations to bi-relations. 
\vspace{-5pt}  

\df
A bi-relation $(R_H, R_S)$ on a topological space is continuous if $R_H$ has closed sections and $P_S$ has open sections.
\dff

\vspace{-10pt}  

\df
A bi-relation $(R_H, R_S)$ on a  mixture set $\CS$ is 
 scalarly continuous if $R_H$ is mixture continuous and for all $x,y,z\in \CS$, the sets  
 $\{\lambda \in [0,1] ~|~ x\lambda y\in P_S(z)\} \text{ and } \{\lambda \in [0,1] ~|~ x\lambda y\in P_S^{-1}(z)\}$ are open.  
\lb{df: properties}
\dff 
\vspace{-5pt}  


\nt The second part of scalar continuity  
 is  a property slightly stronger than  $R_S$ being Archimedean. The Archimedean property and part (ii) of Definition \ref{df: properties} are equivalent if $R_S$ is mixture-continuous; see   \citet*[Proposition 1]{gku19} for details. We leave it to the reader to  obtain the versions of the two propositions above for bi-relations.



\section{Some Concluding Examples}

We conclude the paper with three examples. The first is a modified version of a famous example due to  \citet*{gp84}, and illustrates a preference relation that is  $p$-continuous but not continuous, thus ensuring that the two theorems reported in this paper  are non-vacuous. The second example concerns  the canonical order in $\Re^n$; it satisfies all of the assumptions of Theorem \ref{thm: bict} except  the openness requirements of the sections. The third example  is a simple modification of the second that is phrased in terms of a relation that is  both incomplete and non-transitive.

\medskip

\nt {\bf Example 1.}  Let $X=[0,1]^2$ and $f:X\ra \Re$ defined as 
$$
f(x)=\frac{2x_1x_2}{x_1^2+x_2^2} \text{ if } x\neq (0,0) \text{ and } f(0,0)=(1,1).
$$
Note that $f$ is not continuous in each variable, and hence not (jointly) continuous. Induce a relation  $R$  on $X$ such that  $(x,y)\in R$ if $f(x)\leq f(y)$. It is clear that $R$ is complete and transitive. However, $R$ is not mixture-continuous, hence not continuous. 
  In order to see this, let $x= (0,0)$ and $y=(1,0)$.  Then  the set $\{\lambda\in [0,1]|\lambda x + (1-\lambda) y\in R^{-1}(y)\}=(0,1]$ is not open in $X$. It is not difficult to show that $R$ is $p$-continuous; simply for  $x,y$ above, set $\Lambda_{xy}=[0,1]$ and $f_{xy}(\lambda)=(\lambda, \lambda)$ for $\lambda<0.5$ and $f_{xy}(\lambda)=(\lambda, 1-\lambda)$ for $\lambda\geq 0.5$.  \qed

\medskip

\nt {\bf Example 2.} Let $X=\Re_+^n$ and $R$ be the usual relation $\geq$ on $X$ defined as ``$x\geq y$ if $x_i\geq y_i$ for all $i$.'' The asymmetric part $P$ of $R$ is the relation $>$ defined as ``$x>y$ if $x_i\geq y_i$ for all $i$ and $x_j>y_j$ for some $j$.'' It is clear that $X$ is connected,  $R$ is anti-symmetric, transitive, non-trivial and has closed sections. However the section of $P$ are not open since $P(x)=R(x)\backslash \{x\}$  and $P^{-1}(x)=R^{-1}(x)\backslash \{x\}$. Clearly, $R$ is incomplete. \qed 

\medskip

\nt {\bf  Example 3.}  Let $X=[0,1]$ and $R=\{(x,y)|x\leq 0.5, x\leq y\leq 0.5\}\cup \{(x,y)|x\geq 0.5, y\geq x\}$. It is clear that $R$ is reflexive and anti-symmetric, hence $R$ is semi-transitive and has a transitive indifference. Note that $R^{-1}(0.5)=[0, 0.5]$ and $R(0.5)=[0.5, 1]$. Since $[0, 0.5]\times [0.5, 1]\nsubseteq R$ implies $R$ is not transitive. 
 It is clear that $R$ is not complete. Since $R$ has closed graph, it has closed sections. However, $P$ does not have open sections. For example $P(0.25)=(0.25, 0.5]$ and $P^{-1}(0.75)=[0.5, 0.75)$ which are not open in [0,1].     \qed


\bigskip

\setlength{\bibsep}{5pt}
\setstretch{1.05}


\bibliographystyle{econ} 
\bibliography{References.bib}

\end{document}